\documentclass[prd,showpacs,amsmath,amssymb,superscriptaddress,nofootinbib,showkeys]{revtex4}
\usepackage{amssymb}
\usepackage{amsmath}
\usepackage[cp1251]{inputenc}
\usepackage[T2A]{fontenc}
\usepackage{graphicx}
\usepackage{floatflt}

\begin{document}
\title{Exact cosmological solutions with magnetic field in the theory of gravity with
non-minimal kinetic coupling}

\author{Ruslan K. Muharlyamov}
\email{rmukhar@mail.ru} \affiliation{Department of General
Relativity and Gravitation, Institute of Physics, Kazan Federal
University, Kremlevskaya str. 18, Kazan 420008, Russia}

\author{Tatiana N. Pankratyeva}
\email{ghjkl.15@list.ru} \affiliation{Department of Higher
Mathematics, Kazan State Power Engineering University,
Krasnoselskaya str. 51, Kazan 420066, Russia}

\author{Shehabaldeen O.A. Bashir}
\email{shehapbashir@gmail.com} \affiliation{Department of General
Relativity and Gravitation, Institute of Physics, Kazan Federal
University, Kremlevskaya str. 18, Kazan 420008, Russia; Department of Physics, Faculty of Science, University of Khartoum, Khartoum 11115, Sudan}

\begin{abstract}
We investigate anisotropic and homogeneous cosmological models in the scalar-tensor theory of gravity with non-minimal kinetic coupling of a scalar field to the curvature given by the function $\eta\cdot(\phi/2)\cdot G_{\mu\nu}\,\nabla^\mu \nabla^\nu
\phi$ in the Lagrangian. We assume that the space-times are filled a global unidirectional magnetic field that minimally interacts with the scalar field. The matter sector is not included, since the model is studied in relation to the early times of the Universe evolution. We limit ourselves to the period before and during primary inflation.  The Horndeski theory allows anisotropy to grow over time. The question arises about isotropization. In the theory under consideration, a zero scalar charge imposes a condition on the anisotropy level, namely its dynamics develops in a limited region. This condition uniquely determines a viable branch of solutions of the field equations. The magnetic energy density that corresponds to this branch is a bounded function of time. The sign of parameter $l=1+\varepsilon\eta \Lambda/\mu$ determines the properties of  cosmological models, where $\Lambda$ is the cosmological constant,  $\mu=M^2_{PL}$ is the Planck mass squared. The sign $\varepsilon=\pm 1$ defines the canonical scalar field and the phantom field, respectively.
An inequality $\varepsilon/\eta>0$ is a necessary condition for isotropization of models, but not sufficient.  The model with $l>0$ has the necessary properties: isotropization during expansion, rapid transition to inflationary expansion ($a(t)\propto e^{\sqrt{\frac{\varepsilon}{3\eta}}\cdot t}$), absence of ghost and Laplace instabilities. In other cases  $l\ngtr 0$, the model has various disadvantages. Constraints on the tensor-to-scalar ratio,  the conditions for avoidance of ghost and Laplacian instabilities lead to the inequalities: $\Lambda>0$, $\eta>0$, $\varepsilon=1$, $1<\Lambda\eta/\mu<1.049$.

\end{abstract}



\maketitle

\section{Introduction}

The rapid growth of the observational cosmology have essentially expanded our knowledge about the Universe \cite{Komatsu,Susuki,Hinshaw,PAR,DESI1,DESI2}. The discovery of the accelerated expansion of the Universe and the dark matter evidence motivates the development of modified theories of gravity. A common version of the extension of gravity theories  is   the Horndeski gravity (HG) \cite{Horndeski}. For the HG, the equations of motion have derivative order no higher than second. Within this criterion, the HG is the most general and interesting extension of the scalar tensor theory of gravitation.

In this paper we study cosmological models with a magnetic field
in the subclass of HG with a non-minimal kinetic coupling (NMKC) of a scalar field $\phi$ with the Einstein tensor with the action density
\begin{eqnarray}\label{lagr1} L_H=\sqrt{-g}\left(\frac{\mu R}{2}+G_2(\phi,X)+\frac{\eta\phi}{2}G_{\mu\nu}\,\nabla^\mu \nabla^\nu
\phi\right) \,,\end{eqnarray}
where $g$ is the determinant of metric tensor
$g_{\mu\nu}$; $R$ is the Ricci scalar and $G_{\mu\nu}$ is the
Einstein tensor;  $X=-\frac{1}{2}\nabla^\mu\phi \nabla_\mu\phi$ is the kinetic
term;  $\mu=M^2_{PL}$ is the Planck mass squared.   NMKC involves the additional dimensional parameter $\eta$ with dimension of {\it (length)$^2$}, $[\eta]=[L^2]=[T^2]=[M^{-2}]$. Here we assume $c=\hbar=1$.
We choose the electromagnetic part in the form
\begin{equation}\label{lagrF}L_{F}=-\frac{\sqrt{-g}}{4}F_{\mu\nu}F^{\mu\nu},\end{equation}
where $F_{\mu\nu}$ is the electromagnetic field. The matter sector is not included, since the model is studied in relation to the early times of the Universe evolution.

NMKC reveals various interesting features of astrophysical objects: wormholes \cite{Sushkov,Korolev}, black holes \cite{Rinaldi,Minamitsuji,Kobayashibb,Babichev}, neutron stars \cite{Cisterna,Maselli,Silva,Kashargin}.
In works \cite{Sush2009,Sar2010,Sush2012,Sush2020,SkuSusTop2013,MatSus2015,StaSusVol2016,StaSusVol2019} present interesting cosmological models. In these models, inflation occurs through a new mechanism without tuned the potential. A realistic cosmological model must describe several different phases of the Universe evolution  (the primary inflation, the matter-dominated stage, the present acceleration). The influence of the NMKC decreases over time and thereby a change in cosmological epochs occurs. These properties are manifestations of the screening mechanism characteristic of the NMKC. At early times of the Universe evolution, the $\Lambda$-term and matter are screened. In work \cite{StaSusVol2019},  it was found that anisotropy is screened at early time within the Bianchi I  space-time model (BI).

As is known, there is a large-scale magnetic field in the Universe. The Universe with this field has been studied by many authors \cite{Doroshkevich,Thorne,Jacobs,Salimyx,Horwood,Bronnikov,Watanabe,Soda,Do,Nguyen,Petriakova,Muharlyamov24,Muharlyamov241}.
Various hypotheses are proposed about the origin of the magnetic component of the Universe. Here we agree with the assumption of a primary origin of the magnetic field, i.e the field arises at the very initial stages of the Universe evolution. The scalar field is the cause of the primary inflation of the Universe, so it is interesting to study the interaction of the magnetic field with the scalar field. The combined influence of these fields on the Universe development expands the possibilities of observing the dark sector of the Universe.

We will consider the global magnetic field within the framework of the BI. The BI models are of great interest to researchers  \cite{Muharlyamov0,SushkovStar,Muharlyamov2,Muharlyamov3,Koussour,Koussour1,Akarsu,Sarmah,Hawking0,Hu,Momeni,Momeni1,Mehran,Marco}. Planck’s data on the temperature and the polarization of the cosmic microwave background (CMB) radiation allow  researchers to obtain constraints on the isotropy of the Universe, for example, in work \cite{Saadeh}. Anomalies were noticed on large scales of the CMB radiation. The works \cite{Komatsu,Ellis, Thorsrud} show that these anomalies can be explained within the Bianchi Universe. These considerations do not exclude the possibility of the existence of an anisotropic early Universe.

In this work we explore in details cosmological solutions in the theory (\ref{lagr1}), (\ref{lagrF}) with
\begin{equation} \label{g02} G_2=\varepsilon X-\Lambda,\,\,  \varepsilon = \pm1,\end{equation}
where $\Lambda$ is the cosmological constant  with the dimension $[\Lambda]=[L^{-4}]$. The sign $\varepsilon=\pm 1$ defines the canonical scalar field and the phantom field, respectively. We consider the issue of isotropization of the model. We study the combined influence of the NMKC and the magnetic field on the anisotropic properties of space-time  at the early times of the Universe. We limit ourselves to the period before and during primary inflation.

\section{Field equations}\label{sec2}

The homogeneous and anisotropic Bianchi I metric has the form
\begin{equation}
 ds^2 = -dt^2 + a^2_1(t)dx_1^2 + a^2_2(t)dx_2^2 + a^2_3(t)dx_3^2. \label{met0}
\end{equation}

Then we have the gravity equations:
\begin{eqnarray} \label{00}
 G^0_0\left(\mu+\frac{3\eta \dot{\phi}^2}{2}\right) = G_2 -
G_{2X}\dot{\phi}^2+T^{({\rm em})0}_{0}\,,\end{eqnarray}
\begin{eqnarray} \label{ii}G^{i}_{i}\left(\mu+\frac{\eta
\dot{\phi}^2}{2}\right)-(H_{j}+H_{k})\frac{d}{dt}\left(\mu+\frac{\eta
\dot{\phi}^2}{2}\right) =G_2 +T^{({\rm
em})i}_{i} \,.
\end{eqnarray}
Here the dot denotes  the $t$-derivative, one has $H_i=\dot {
a}_i/{ a}_i$, and the average Hubble parameter is
$H=\dfrac{1}{3}\sum\limits_{i=1}^3 H_i\equiv \dot{ a}/{ a}$ with
${ a}=({ a}_1{ a}_2{ a}_3)^{1/3}$ -- the geometric mean  scale factor; $T^{({\rm em})\mu}_{\nu}$ -- the stress–energy tensor of the electromagnetic field.  The Einstein tensor components
are
\begin{eqnarray}
&&G^0_0=-\left(H_1H_2 +H_2H_3 +H_3H_1\right)\,, \\
&&G^{i}_{i}=-\left(\dot{H}_{j} +\dot{H}_{k} +H_{j}^2 +H_{k}^2
+H_{j}H_{k}\right)\,,
\end{eqnarray}
where the triples of indices $\{i,j,k\}$ take values $\{1,2,3\}$,
$\{2,3,1\}$, or $\{3,1,2\}$.

The equation for the scalar field $\phi(t)$  can be represented as
\begin{eqnarray}\label{scalar}
\frac{1}{a^3}\frac{d}{dt}\left({ a^3}\dot{\phi}\, \Big[ G_{2X}+\eta G^0_0
\Big]\right)=G_{2\phi}.
\end{eqnarray}
Electromagnetic equations have the form $\partial_\mu[a^3F^{\mu\nu}]=0$ and there is the Bianchi identity $\nabla_\mu F_{\nu\alpha}+\nabla_\alpha F_ {\mu\nu}+\nabla_\nu F_{\alpha\mu}=0$.
We assume that there is the magnetic field
having the same direction  $x_3$.
The magnetic field corresponds to non-zero components $F_{\gamma\delta}$: $
F_{21}=-F_{12}=q_m$, where $q_m$ is constant.
The magnetic field strength is determined by the equality
\begin{eqnarray}\label{mag stren}
B^2=F_{21}F^{21}=\frac{q^2_m}{a^2_1a^2_2}\,.
\end{eqnarray}

The stress–energy tensor of the electromagnetic field  is written like this:
\begin{equation}
T^{({\rm em})\mu}_{\nu}= -\frac{1}{4}\delta^\mu_\nu
F_{\gamma\delta}F^{\gamma\delta} + F_{\nu \beta} F^{\mu
\beta} \,. \label{em1}
\end{equation}
The tensor $T^{({\rm em})\mu}_{\nu}$ has non-zero components:
\begin{eqnarray}\label{tem}T^{({\rm em})0}_{0}=T^{({\rm em})3}_{3}=-T^{({\rm em})1}_{1}=-T^{({\rm em})2}_{2}=-\mathcal{E}^{({\rm em})}\,,\end{eqnarray}
where $\mathcal{E}^{({\rm em})}$ -- the magnetic field energy density:
\begin{equation}\label{eem}\mathcal{E}^{({\rm em})}=\frac{1}{2}B^2=\frac{q_m^2}{2a^2_1a^2_2}.\end{equation}

Let's consider a special case $a_1=a_2$  that corresponds the locally rotationally symmetric
(LRS) Bianchi-I space-time. We will take the following parameterization the scale factors:
\begin{equation}
ds^2 =
-dt^2+e^{2\alpha(t)}[e^{2\beta(t)}(dx^2+dy^2)+e^{-4\beta(t)}dz^2].
\label{met}
\end{equation}
Then the Hubble parameters are given by
\begin{equation}\label{H}
H_1=H_2=\dot{\alpha}+\dot{\beta}\,,\,
H_3=\dot{\alpha}-2\dot{\beta}\,,\,H=\dot{\alpha}\,\, \, (a=e^{\alpha(t)})\,.
\end{equation}
The function $e^{\beta}$ represents the deviation from isotropy,
while $e^{\alpha}$ is the isotropic part.

We choose function $G_2$ in the form
\begin{equation} \label{g2} G_2=\varepsilon X-\Lambda, \end{equation}
where $\Lambda$ is the cosmological constant. The parameter $\varepsilon$ takes the value $+1$ for the canonical scalar
field and $-1$ for the phantom one.

In view of (\ref{tem}), (\ref{eem}), (\ref{met}), (\ref{H}) and (\ref{g2}), the system of field equations (\ref{00}), (\ref{ii}), (\ref{scalar}) has the consequences
\begin{equation} \label{lr00}3\big(\dot{\alpha}^2-\dot{\beta}^2\big)\left(\mu+\frac{3\eta \dot{\phi}^2}{2}\right)
=\frac{q^2_m}{2}\cdot e^{-4(\alpha+\beta)}
+\varepsilon\frac{\dot{\phi}^2}{2}+\Lambda,\end{equation}
$$\left(\mu+\frac{\eta
\dot{\phi}^2}{2}\right)\big(2\ddot{\alpha}+3\dot{\alpha}^2 +
3\dot{\beta}^2\big)+2\dot{\alpha}\frac{d}{dt}\left(\mu+\frac{\eta
\dot{\phi}^2}{2}\right)=$$\begin{equation} \label{lrii}=-\frac{q^2_m}{6}\cdot e^{-4(\alpha+\beta)}-\varepsilon\frac{\dot{\phi}^2}{2}+\Lambda,\end{equation}
\begin{equation} \label{lrb}e^{-3\alpha}\cdot\frac{d}{dt}\left[\left(\mu+\frac{\eta
\dot{\phi}^2}{2}\right)e^{3\alpha}\dot{\beta}\right]=\frac{q^2_m}{3}\cdot e^{-4(\alpha+\beta)},\end{equation}
\begin{eqnarray}\label{scalar1}
\dot{\phi}\, \Big[ \varepsilon
-3\eta \big(\dot{\alpha}^2-\dot{\beta}^2\big)\Big]=C_{\phi}e^{-3\alpha}\,,\end{eqnarray}
where $C_{\phi}$ is the scalar charge.
The magnetic field energy density is
\begin{equation}\mathcal{E}^{({\rm em})}=\frac{q^2_m}{2}\cdot e^{-4(\alpha+\beta)}.\end{equation}
The system (\ref{lr00})-(\ref{scalar1}) contains three independent equations.

Further, we put $C_{\phi}=0$.  In this case, there are two branches of the solution.

One branch is defined by the equations $\dot{\phi}=0$ and
\begin{equation}3\mu\big(\dot{\alpha}^2-\dot{\beta}^2\big)=\frac{q^2_m}{2}\cdot e^{-4(\alpha+\beta)}
+\Lambda,\,\, \mu e^{-3\alpha}\cdot\frac{d[e^{3\alpha}\dot{\beta}]}{dt}=\frac{q^2_m}{3}\cdot e^{-4(\alpha+\beta)}.\end{equation}
This branch corresponds to General Relativity. At late time, the Universe expansion is driven by $\Lambda$.

The second branch is determined by another consequence from (\ref{scalar1}):
\begin{equation}\label{dotbeta}\dot{\alpha}^2=\dot{\beta}^2+\frac{\varepsilon}{3\eta}.\end{equation}
Further we will consider only this branch. In isotropic space-time ($\dot{\beta}=q_m=0$) the Universe accelerates according to the law de Sitter or anti-de Sitter with parameters $\dot{\alpha}=\pm\sqrt{\frac{\varepsilon}{3\eta}}$.  $\Lambda$-term is screened and makes no contribution to the Universe acceleration. In the anisotropic case ($\dot{\beta}\neq 0$) the question of screening remains open.
Nevertheless,  it can be said that in the presence of $\dot{\beta}\rightarrow 0$,  the geometric mean behavior of the Universe is determined by parameter $\sqrt{\frac{\varepsilon}{3\eta}}$:  $a(t)\propto e^{\pm\sqrt{\frac{\varepsilon}{3\eta}}\cdot t}$.

We will use the following criterion for isotropization of the cosmological model:
\begin{equation}\label{critizotrop} \frac{\dot{\beta}^2}{\dot{\alpha}^2}=\beta'^2_\alpha \rightarrow 0 \,\,\, \text{as} \,\,\, \alpha \rightarrow +\infty.\end{equation}
Ratio $|\dot{\beta}/\dot{\alpha}|=|\beta'_\alpha|$ determines the  anisotropy level. We will need another form of (\ref{dotbeta}):
\begin{equation}\label{dbdf}\frac{\dot{\beta}^2}{\dot{\alpha}^2}=\beta'^2_\alpha=1-\frac{\varepsilon}{3\eta}\cdot\frac{1}{\dot{\alpha}^2}.\end{equation}
A necessary but not sufficient condition for the fulfillment of (\ref{critizotrop}) is the inequality
\begin{equation}\label{ettavar}\frac{\varepsilon}{\eta}>0.\end{equation}
Parameters $\varepsilon$ and $\eta$ have the same sign. Let's enter the parameter
\begin{equation}\label{hett12} h_\eta=\sqrt{\frac{\varepsilon}{3\eta}}.\end{equation}
Equality (\ref{dbdf}) can be rewritten as follows
\begin{equation}\label{alf}\dot{\alpha}^2=\frac{h_\eta^2}{1-\beta'^2_\alpha}.\end{equation}
From here we get obvious restrictions:
\begin{equation}\label{bnaal}|\beta'_\alpha|<1,\end{equation}
\begin{equation}\label{h1}|H|=|\dot{\alpha}|\geq h_\eta.\end{equation}
The zero charge $C_\phi$ leads to the requirement that the anisotropy level $\beta'_\alpha$ be limited and that there is a lower non-zero limit to the Hubble parameter $|H|$.
The isotropization condition (\ref{critizotrop}) will be satisfied if $\dot{\alpha}^2$ is a decreasing function at $t\rightarrow +\infty$.

Substituting (\ref{dotbeta}) into equation (\ref{lr00}) gives
\begin{equation}\label{phem}\dot{\phi}^2=\varepsilon\mu\left[3(\varepsilon_\Lambda h_\Lambda^2-h_\eta^2)+\frac{q^2_m}{2\mu}\cdot e^{-4(\alpha+\beta)}\right],\end{equation}
where $h_\Lambda=\sqrt{|\Lambda|/(3\mu)}$, $\varepsilon_\Lambda=\text{sign}(\Lambda)$.
Parameters $h_\eta$, $h_\Lambda$ have the dimension of the Hubble parameter: $[h_\eta]=[h_\Lambda]=[H]=[T^{-1}]=[L^{-1}]$, and
$[q_m]=[M^2]=[L^{-2}]$. Based on (\ref{phem}) ($\dot{\phi}^2\geq0$), it can be stated that not all values of parameters
$\varepsilon$, $h_\eta$, $\varepsilon_\Lambda$, $h_\Lambda$ allow the density $\mathcal{E}^{({\rm em})}\rightarrow +\infty , \,\, 0$.

Using equalities (\ref{alf}), (\ref{phem}), the equation (\ref{lrii}) is rewritten as follows
$$\left(l+\frac{q^2_m}{6\mu h_\eta^2}\cdot e^{-4(\alpha+\beta)}\right)
\beta''_{\alpha\alpha}+(\beta'^2_\alpha-1)\times$$
\begin{equation}\label{balal}\times\left[\beta'_\alpha\left(-3l+\frac{q^2_m}{6\mu h_\eta^2}\cdot e^{-4(\alpha+\beta)}\right)+\frac{4 q^2_m}{6\mu h_\eta^2}\cdot e^{-4(\alpha+\beta)}\right]=0,\end{equation}
where
\begin{equation}\label{l} l\equiv 1+\frac{\varepsilon\eta \Lambda}{\mu}= 1+\varepsilon_\Lambda\cdot\left(\frac{h_\Lambda}{h_\eta}\right)^2.\end{equation}

Next, we will analyze the properties of the model depending on the numerical parameters.

\section{Model without magnetic field}

Here we will talk about the influence of NMKC without the magnetic field on the space-time. In equation (\ref{balal}) we put $q_m=0$, then
\begin{equation}\label{bezmagn}\beta''_{\alpha\alpha}-3\beta'_\alpha(\beta'^2_\alpha-1)=0.\end{equation}
This equation has several solutions.

{\it 1.} $\beta'_\alpha=\pm 1 \Rightarrow \beta=\pm \alpha+\text{const}$. This is the solution does not fall into region $|\beta'_\alpha|<1$.

{\it 2.} $\beta'_\alpha=0 \Rightarrow \beta=\text{const}$. This is Friedman's flat world without anisotropy that has accelerated expansion according to  the de Sitter's law:
\begin{equation}\label{00sdf}\alpha(t)=h_\eta\cdot t,\,\,\text{or}\,\, a(t)=a_0e^{h_\eta\cdot t}\,.\end{equation}
The scalar field depends linearly on time:
\begin{equation}\label{0sdf}\phi(t)=\pm\left[3\varepsilon\mu (\varepsilon_\Lambda h_\Lambda^2-h_\eta^2)\right]^{1/2}\cdot t\,,\end{equation}
and we have a limitation
\begin{equation}\label{0sdf11}\varepsilon(\varepsilon_\Lambda h_\Lambda^2-h_\eta^2)>0.\end{equation}

{\it 3.} Next corollary (\ref{bezmagn}):
\begin{equation}\label{bezmalf}\beta'^2_\alpha=\frac{1}{1+\text{const}\cdot e^{6\alpha}}.\end{equation}
From (\ref{alf}), (\ref{bezmalf}) it follows
\begin{equation}\label{albet1}\dot{\alpha}^2=h_\eta^2+\sigma_0^2\cdot\frac{a_0^6}{e^{6\alpha}},\,\, \dot{\beta}=\sigma_0\cdot\frac{a_0^3}{e^{3\alpha}},\,\, \sigma_0\lessgtr 0.\end{equation}
The model has an initial singularity at $t\rightarrow 0$: \begin{equation}\label{paaq}a\propto (h_\eta\cdot t)^{1/3} \rightarrow 0, \,\, \dot{\alpha}\propto 1/t\rightarrow\infty.\end{equation}
According to (\ref{paaq}), the Universe begins its expansion without acceleration.
The isotropization condition is satisfied: $|\beta'_\alpha|\propto e^{-3h_\eta \cdot t}\rightarrow 0$, $h_\eta\cdot t\gg 1$, and
the metric potential $a(t)$ approaches (\ref{00sdf}). Consequently, at a certain point in time a phase of accelerated expansion occurs.
The scalar field $\phi(t)$ has the form (\ref{0sdf}) for any $t$.

The model (\ref{albet1}) is similar to the well-known with $L=\sqrt{-g}\left(\mu R/2-\mathbf{\Lambda}\right)$ in the Bianchi I metric, where $\mathbf{\Lambda}=\varepsilon\mu/\eta>0$. There is an exact solution
$$a(t)=e^{\alpha}=\frac{a_0|\sigma_0|^{1/3}}{h_\eta^{1/3}}\cdot \sinh^{1/3}(3h_\eta\cdot t),$$\begin{equation}
e^\beta=b_{1}\cdot[\tanh(3h_\eta\cdot t/2)]^{\frac{1}{3}\cdot \text{sign}(\sigma_0)},\,\, t\geq 0.\end{equation}
The scale factor $a(t)$ is a monotonic increasing function, i.e. the Universe is expanding all the time.
Scale factors $a_i$ have the form:
$$a_{1,2}=\frac{a_0b_1|\sigma_0|^{1/3}}{h_\eta^{1/3}}\cdot \{\sinh(3h_\eta\cdot t)[\tanh(3h_\eta\cdot t/2)]^{\text{sign}(\sigma_0)}\}^{1/3},$$
\begin{equation}\label{bnv123}a_3=\frac{a_0|\sigma_0|^{1/3}}{b_1^{2}h_\eta^{1/3}}\cdot\left\{\frac{\sinh(3h_\eta\cdot t)}{[\tanh(3h_\eta\cdot t/2)]^{2\text{sign}(\sigma_0)}}\right\}^{1/3}.\end{equation}
From here follow the limits $a_i/a=\text{const}_i\neq 0$, $h_\eta\cdot t\gg 1 $. This confirms the isotropization process.

Thus, we see that NMKC without magnetic field does not interfere with the process of the Universe isotropization. After a post-singularity era the Universe enters a primary quasi-de Sitter epoch with the parameter $h_\eta$. Metric functions of isotropic model (\ref{00sdf}) and anisotropic model (\ref{bnv123}) do not contain the $\Lambda$-term; it is replaced by $\eta$: $|\Lambda|/\mu\rightarrow \varepsilon/\eta$. In other words, the $\Lambda$-term is screened. The driving force of primary inflation is  NMKC through parameter $\eta$. Only the scalar field "feels"\, the $\Lambda$-term (see (\ref{0sdf})). The $\Lambda$-term is not clearly present in the space-time dynamics, but, firstly, it influences the character of the scalar field. For example,  if $h_\Lambda=0$ ($\Lambda=0$) or $\varepsilon_\Lambda=-1$ ($\Lambda<0$), only the phantom scalar field ($\varepsilon=-1$) is allowed (see (\ref{0sdf11})) and therefore $\eta<0$ (see (\ref{ettavar})). Secondly, it will be shown below that the $\Lambda$-term is an important factor in the conditions for avoidance of ghost and Laplacian instabilities of models. Next we will study the combined influence of NMKC and the magnetic field.

\section{Models with $l=0$}

Value $l=0$ corresponds to $\Lambda<0$, $h_\eta=h_\Lambda$.
Equation (\ref{balal}) becomes
\begin{equation}\label{betlzero}\beta''_{\alpha\alpha}+(\beta'^2_\alpha-1)(\beta'_\alpha+4)=0.\end{equation}
This equation has several solutions. Solutions $\beta=\pm\alpha+c$, $\beta=-4\alpha+c$ are not in the scope $|\beta'_\alpha|<1$.
Equation (\ref{betlzero}) gives the corollary
\begin{equation}\label{fdg1}\frac{(1-\beta'_\alpha)^3(\beta'_\alpha+4)^2}{(1+\beta'_\alpha)^5}=\text{const}\cdot e^{-30\alpha}.\end{equation}
 Equation (\ref{fdg1}) does not contain a solution with isotropization: $\beta'_\alpha\rightarrow 1,\,\,-4\neq0\,\, \text{as} \,\, \alpha\rightarrow +\infty.$

Thus, NMKC with the magnetic field and parameter $l=0$ blocks the isotropization process. From this point of view, the model is not of interest.

\section{Solving field equations with the parameter $l\neq0$}

Here we will present the solution of field equations with the parameter $l\neq0$.
In equation (\ref{balal}) we will make a replacement for the unknown function $\beta(\alpha)$:
\begin{equation}\label{ual0}u(\alpha)=\beta(\alpha)+\alpha,\end{equation}
then
\begin{equation}\label{ualal}u''_{\alpha\alpha}+u'_\alpha(u'_\alpha-2)\left[3+u'_\alpha\cdot \frac{-3l+\frac{q^2_m}{6\mu h_\eta^2}\cdot e^{-u}}{l+\frac{ q^2_m}{6\mu h_\eta^2}\cdot e^{-u}}\right]=0.\end{equation}
The requirement (\ref{bnaal}) leads to
\begin{equation}\label{ineq} 0<u_\alpha<2.\end{equation}
These inequalities cut off unnecessary solutions of the equation (\ref{ualal}). The equation  gives
$$u'_\alpha=2+\left(l+\frac{ q^2_m}{6\mu h_\eta^2}\cdot e^{-4u}\right)^2\times$$
\begin{equation}\label{ualal1}\times\left(c_0e^{-3u}-l^2+2l\cdot \frac{ q^2_m}{6\mu h_\eta^2}e^{-4u}-\frac{1}{5}\left(\frac{ q^2_m}{6\mu h_\eta^2}\right)^2 e^{-8u}\right)^{-1},\end{equation}
where $c_0$ -- integration constant. We'll put $c_0=0$.

Let's introduce the dimensionless function
\begin{equation}\label{sal0}s=\frac{1}{|l|}\cdot\frac{ q^2_m}{6\mu h_\eta^2}\cdot e^{-4u}>0\end{equation}
that is associated with the magnetic field energy density $\mathcal{E}^{({\rm em})}$:
\begin{equation}s=\frac{1}{3\mu h_\eta^2|l|}\cdot\mathcal{E}^{({\rm em})}>0.\end{equation}
Equation (\ref{ualal1}) will be rewritten
\begin{equation}\label{sal}u'_\alpha=-\frac{s'_\alpha}{4s}=-\frac{3s^2+30\varepsilon_l\cdot s-5}{s^2-10\varepsilon_l\cdot s+5},\end{equation}
where $\varepsilon_l=\text{sign} (l)$. Next we need a representation:
\begin{equation}3s^2+30\varepsilon_l\cdot s-5=3(s-s_1)(s-s_2),\end{equation}
\begin{equation}s^2-10\varepsilon_l\cdot s+5=(s-s_3)(s-s_4),\end{equation}
where
\begin{equation}s_1=-5\varepsilon_l-4\cdot \sqrt{\frac{5}{3}}<0,\,\, s_2=-5\varepsilon_l+4\cdot \sqrt{\frac{5}{3}}>0,\end{equation}
\begin{equation}s_3=5\varepsilon_l-2\sqrt{5},\,\, s_4=5\varepsilon_l+2\sqrt{5}.\end{equation}

Inequalities (\ref{ineq}) lead to restrictions
\begin{equation}\label{s2}\varepsilon_l=1:\,\,0<s<s_2,\end{equation}
\begin{equation}\label{s21}\varepsilon_l=-1:\,\, s\in (0,1)\cup(1,s_2).\end{equation}
That is, within the framework of NMKC, a requirement arises for the finiteness of the magnetic energy density $\mathcal{E}^{({\rm em})}$. In General Relativity, within the framework of cosmology, density has a singularity: $\mathcal{E}^{({\rm em})}\rightarrow +\infty$ as $a\rightarrow 0$.

Integrating equation (\ref{sal}), we get
\begin{equation}\label{als}e^{4\alpha}=a^4_0\cdot\frac{[(s+|s_1|)(s_2-s)]^{2/3}}{s}\end{equation}
or
\begin{equation}\label{albetta}e^{4\beta}=\frac{1}{a^4_0}\cdot\frac{ q^2_m}{6\mu h_\eta^2|l|}\cdot\frac{1}{[(s+|s_1|)(s_2-s)]^{2/3}}.\end{equation}
Scale factors $a_i$ have the form:
$$a_1=a_2=e^{\alpha+\beta}=\left(\frac{ q^2_m}{6\mu h_\eta^2|l|}\right)^{1/4}\cdot s^{-1/4},$$
\begin{equation}\label{ai}a_3=e^{\alpha-2\beta}=a_0^3\cdot\left(\frac{6\mu h_\eta^2|l|}{ q^2_m}\right)^{1/2}\cdot\frac{[(s+|s_1|)(s_2-s)]^{1/2}}{s^{1/4}}.\end{equation}
Equality (\ref{als}) implicitly defines a multivalued function $s=f_i(\alpha)$.
Conditions (\ref{s2}), (\ref{s21})  select an one-valued branch $s(\alpha)$.
Knowing connections (\ref{ual0}), (\ref{sal0}) we can determine $\beta(\alpha)$ from $s(\alpha)$.

 If we find equality $f(s,t)=0$, then we obtain a solution to the system in parametric form with parameter $s$. In view of (\ref{sal}), from equation (\ref{alf}) it follows
$$\dot{\alpha}^2=h_\eta^2\cdot\frac{(s^2-10\varepsilon_l\cdot s+5)^2}{5(s+\varepsilon_l)^2(5-30\varepsilon_l\cdot s-3s^2)}=$$
\begin{equation}\label{dals}=h_\eta^2\cdot\frac{[(s-s_3)(s-s_4)]^2}{15(s+\varepsilon_l)^2(s+|s_1|)(s_2-s)}.\end{equation}
Using identity $\dot{\alpha}=\dot{s}/s'_\alpha$, we get
\begin{equation}\label{dots}\dot{s}^2=h_\eta^2\cdot\frac{16s^2(5-30\varepsilon_l\cdot s-3s^2)}{5(s+\varepsilon_l)^2}.\end{equation}
Then
$$-4h_\eta\cdot (\pm t)=\sqrt{\frac{5}{3}}\arcsin\left[\frac{1}{4}\cdot\sqrt{\frac{3}{5}}(5 +\varepsilon_l\cdot s)\right]+\ln s-$$\begin{equation}\label{tst}-
\ln\left[5-15\varepsilon_l \cdot s+\sqrt{5}\sqrt{5-30\varepsilon_l \cdot s-3s^2}\right]+c.\end{equation}

The function $\dot{\phi}^2$ is expressed in terms of $s$:
\begin{equation}\label{phemlp}\dot{\phi}^2=3\varepsilon\mu\left[\varepsilon_\Lambda h_\Lambda^2-h_\eta^2+(h_\eta^2+\varepsilon_\Lambda h_\Lambda^2)\cdot s\right].\end{equation}
From inequality $\dot{\phi}^2\geq 0$ follow restrictions on parameters $\varepsilon$, $h_\Lambda$, $h_\eta$, $\varepsilon_\Lambda$.

Thus, the system of equalities (\ref{als}), (\ref{albetta}), (\ref{tst}), (\ref{phemlp}) will determine the functions
$\alpha(t)$, $\beta(t)$, $s(t)$, $\phi(t)$.
Next we will analyze solutions for different cases $\varepsilon_l=\pm 1$ ($l\gtrless 0$).

\section{Case $l>0$}

Here we will consider case  $\varepsilon_l=1$ ($l>0$), i.e. $h_\eta^2>-\varepsilon_\Lambda h^2_\Lambda$.
Therefore the maximum value $s(t)$ is
$s_2=4\cdot \sqrt{\frac{5}{3}}-5\approx 0.164$
and accordingly for the energy density:
\begin{equation}\label{maxmagn}\mathcal{E}^{({\rm em})}_{upp}=s_2\cdot 3\mu h_\eta^2|l|\approx 0.5\mu h_\eta^2|l|.\end{equation}

\begin{figure}[h]
\includegraphics[width=10cm]{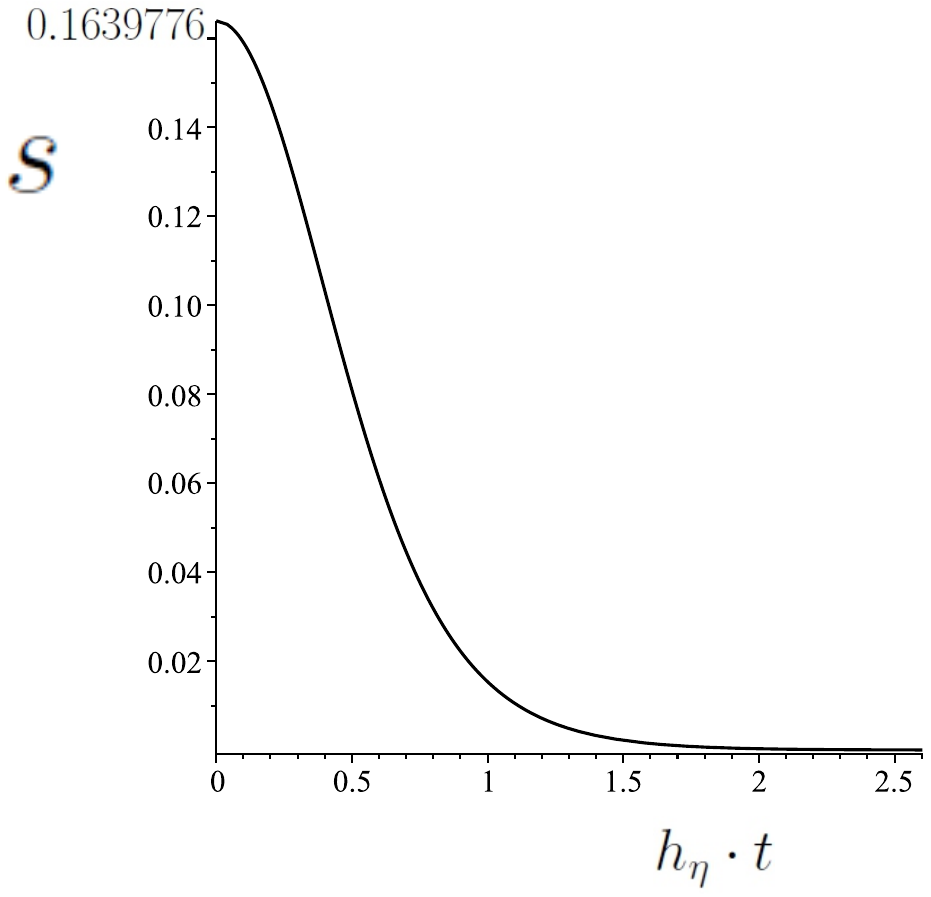}
\caption{The profile  $s(t)$. The dimensionless function $s(t)$ is associated with the magnetic field energy density $\mathcal{E}^{({\rm em})}$: $s=\frac{1}{3\mu h_\eta^2|l|}\cdot\mathcal{E}^{({\rm em})}$. The graph shows that the magnetic energy density is a bounded and monotonic decreasing function. The maximum value $s(t)$ is
$s_2=4\cdot \sqrt{\frac{5}{3}}-5\approx 0.164$. \label{magnplotn}}
\end{figure}
The "+"\, sign in (\ref{tst}) will correspond to the Universe expansion  over time. The choice of the integration constant $c=-\sqrt{\frac{5}{3}}\cdot\frac{\pi}{2}-\ln \frac{\sqrt{15}}{60}$ means the start of the count at the moment $t=0$ and $s(0)=s_2$.
Thus
$$-4h_\eta\cdot t=\sqrt{\frac{5}{3}}\arcsin\left[\frac{1}{4}\cdot\sqrt{\frac{3}{5}}(s+5)\right]+\ln s-$$\begin{equation}\label{st2}-
\ln\left[5-15s+\sqrt{5}\sqrt{5-30s-3s^2}\right]-\sqrt{\frac{5}{3}}\cdot\frac{\pi}{2}-\ln \frac{\sqrt{15}}{60}.\end{equation}
 Fig.\ref{magnplotn} shows that the magnetic energy density is a bounded and monotonic decreasing function. Value range $s\in(0,s_2]$ corresponds to the entire time interval $0\leq t<+\infty$.

\subsection{Space-time properties of the model}

\begin{figure}[h]
\includegraphics[width=10cm]{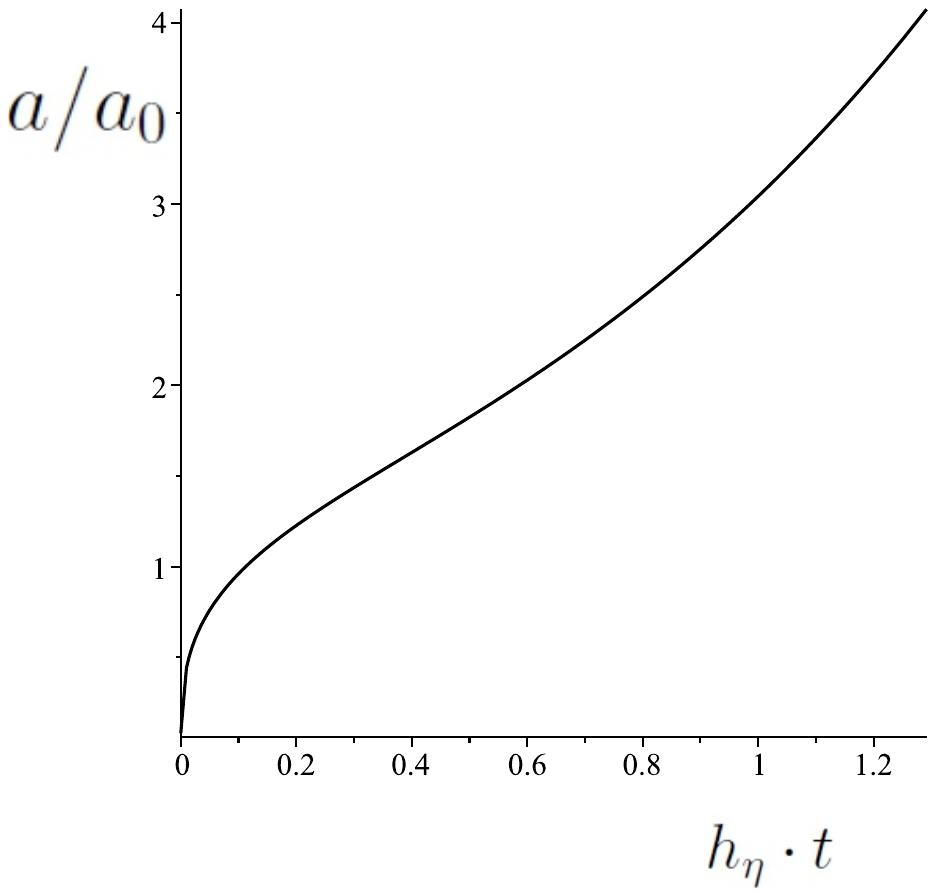}
\caption{The profile  $a/a_0$. The scalar factor  $a(t)=e^{\alpha(t)}$ is an increasing function, i.e. the Universe volume increases with time. The moment $t=0$ corresponds to a cosmological singularity: $a\propto (h_\eta\cdot t)^{1/3} \rightarrow 0$, $\dot{\alpha}\propto 1/t\rightarrow\infty$. \label{a}}
\end{figure}
Using (\ref{als}), (\ref{st2}), we can plot the graph Fig.\ref{a} for the scale factor $a(t)=e^{\alpha(t)}$.
The Universe volume  increases with time. The moment $t=0$ corresponds to a cosmological singularity: $a\propto (h_\eta\cdot t)^{1/3} \rightarrow 0$, $\dot{\alpha}\propto 1/t\rightarrow\infty$ (see (\ref{als}), (\ref{dals})). The Universe begins its expansion without acceleration. Comparing with approximation (\ref{paaq}) without the magnetic field, it is clear that the type of space-time singularity has not changed. However, unlike in General Relativity, the magnetic field does not have a singularity in $t=0$:
\begin{equation}\label{sas1}s\approx s_2[1-3(h_\eta\cdot t)^2],\,\, h_\eta\cdot t\rightarrow 0. \end{equation}
NMKC eliminate such singularity.

In isotropic space, equality $a(t_{init})=0$ means that the Universe expands from a point. In anisotropic space, this condition on the geometric mean scale factor $a=(a_1a_2a_3)^{1/3}$ can be realized through different approximations $a_1$, $a_2$, $a_3$.
 For the model without magnetic field (\ref{bnv123}) there are two possible options. For $\dot{\beta}>0$ ($\sigma_0>0$): $a_{1,2}\propto (h_\eta\cdot t)^{2/3}\rightarrow0$, $a_3\propto (h_\eta\cdot t)^{-1/3}\rightarrow\infty$, i.e. it has a thread-like singularity. For $\dot{\beta}<0$ ($\sigma_0<0$): $a_{1,2}\propto const\neq0\,$, $a_3\propto(h_\eta\cdot t)\rightarrow0$, i.e. it has a pancake singularity. The magnetic field removes uncertainty. In the presented model there is only the last option. Taking into account approximation (\ref{sas1}), from (\ref{ai}) follows: $a_{1,2}\approx\left(\frac{q^2_m}{6\mu h_\eta^2|l|s_2}\right)^{1/4}$, $a_3\approx \left(\frac{6\mu h_\eta^2|l|s_2}{ q^2_m}\right)^{1/2} \cdot b_0\cdot (h_\eta\cdot t)$.

\begin{figure}[h]
\includegraphics[width=10cm]{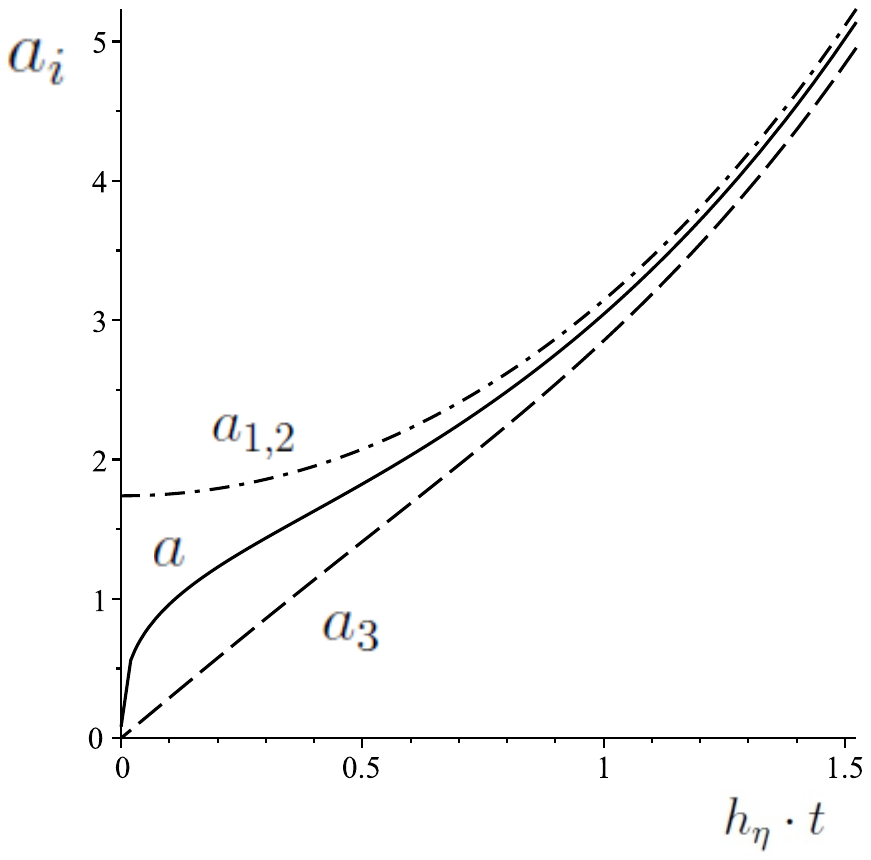}
\caption{The profile  $a_{1,2}$, $a_3$, $a$. The following parameter values are selected here: $a_0=1$, \, $\dfrac{ q^2_m}{6\mu h_\eta^2|l|}=1.5$. The model has  the pancake singularity: $a_{1,2}\approx\left(\frac{q^2_m}{6\mu h_\eta^2|l|s_2}\right)^{1/4}$, $a_3\approx \left(\frac{6\mu h_\eta^2|l|s_2}{ q^2_m}\right)^{1/2} \cdot b_0\cdot (h_\eta\cdot t)$. Scale factors grow monotonically over time, i.e. the Universe is constantly expanding in all directions. \label{a123f}}
\end{figure}

Now let's consider the approximation at a late time. For large $h_\eta\cdot t$, the magnetic field decreases according to the law
\begin{equation}\label{sas2}s\approx s_0 e^{-4h_\eta \cdot t}\sim a^{-4},\,\, h_\eta\cdot t\gg 1.\end{equation}
This approximation is no different from the behavior of the magnetic field within the framework of General Relativity in the process of isotropization. The scale factor $a(t)\propto e^{h_\eta\cdot t}$ corresponds to the phase of accelerated expansion.
Taking into account approximation  (\ref{sas2}), from (\ref{ai}) follows: $a_{1,2}\approx \left(\frac{q^2_m}{6\mu h_\eta^2|l|}\right)^{1/4}\cdot s_0^{-1/4} e^{h_\eta \cdot t}$, $a_3\approx a_0^3\cdot\left(\frac{6\mu h_\eta^2|l|}{ q^2_m}\right)^{1/2}(5/3)^{1/2}\cdot s_0^{-1/4} e^{h_\eta \cdot t}$. Therefore, the process of isotropization occurs: $a_i/a\rightarrow\text{const}_i\neq 0$, $h_\eta \cdot t\rightarrow +\infty$. This conclusion is confirmed by another criterion of isotropization: $|\beta'_\alpha|\rightarrow 0$, $h_\eta\cdot t\rightarrow +\infty$. Let's show it. Using (\ref{ual0}) and (\ref{sal}), we get the anisotropy level
\begin{equation}\dot{\beta}/\dot{\alpha}=\beta'_\alpha=-\frac{4(s^2+5s)}{s^2-10s+5}.\end{equation}
The approximation  (\ref{sas2}) gives the expected result: $|\beta'_\alpha|\propto e^{-4h_\eta \cdot t}\rightarrow 0$, $h_\eta \cdot t\rightarrow +\infty$. For comparison, in the model without the magnetic field, the anisotropy decreases more slowly, $|\beta'_\alpha|\propto e^{-3h_\eta \cdot t}$.
\begin{figure}[h]
\includegraphics[width=10cm]{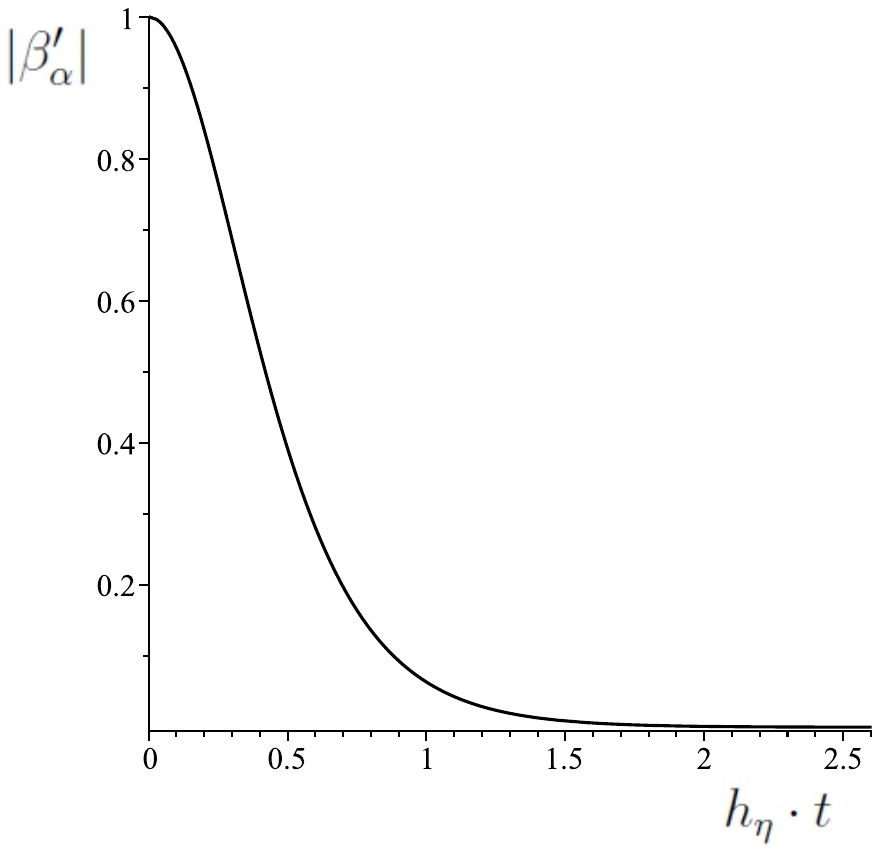}
\caption{The profile  $|\beta'_\alpha|$. The isotropization criterion is satisfied: $|\beta'_\alpha|\propto e^{-4h_\eta \cdot t}\rightarrow 0$, $h_\eta \cdot t\rightarrow +\infty$. Within a few units of $h_\eta \cdot t$, the anisotropy becomes small, therefore $a(t)\propto e^{h_\eta t}$ will be almost throughout the entire inflation time interval. \label{ParamAniz}}
\end{figure}
The behavior $|\beta'_\alpha|$ can be seen in Fig.\ref{ParamAniz}. All values of the quantity $\beta'_\alpha$ fall into region (\ref{bnaal}).  Function $\beta'_\alpha$ has negative values: $-1<\beta'_\alpha<0$, then $\dot{\beta}<0$. This sign of $\dot{\beta}$ leads to $H_{1,2}>0$, $H_3>0$ (see (\ref{H})), i.e. the Universe is constantly expanding in all directions. The conclusion is clearly confirmed in Fig.\ref{a123f}. Scale factors grow monotonically over time.

\subsection{Time frame of the model}

\begin{figure}[h]
\includegraphics[width=10cm]{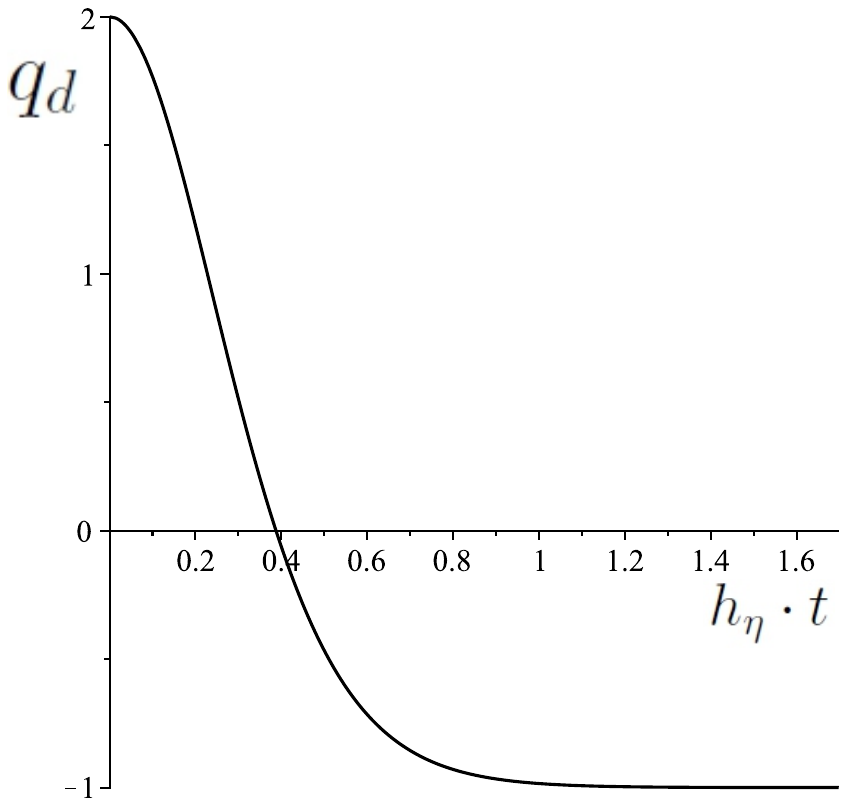}
\caption{The deceleration parameter $q_d$ profile. There are two phases. In the first phase,
there is no acceleration ($q_d\geq0$), and the second one is characterized by the acceleration expansion of the Universe ($q_d<0$).
The beginning of primary inflation corresponds to the value $h_\eta\cdot t_b\approx 0.387$ ($q_d(h_\eta\cdot t_b)=0$).\label{q}}
\end{figure}

Earlier we noted that according to the model the Universe begins to expand without acceleration. At relatively late times there is a phase of accelerated expansion. Complete information about the expansion phases is provided by the deceleration parameter (DP):
\begin{equation}q_d=\frac{d}{dt}\left(\frac{1}{\dot{\alpha}}\right)-1=-\frac{64s^2(3s^2+10s-25)}{(s^2-10s+5)^2(1+s)}-1.\end{equation}
From Fig.\ref{q} it is clear that  DP  evolves from positive values at past epoch to
negative values at late time.
In this model, there are two phases. In the first phase,
there is no acceleration ($q_d\geq0$), and the second one is characterized by the acceleration expansion of the Universe ($q_d<0$).
The beginning of primary inflation corresponds to the value $h_\eta\cdot t_b\approx 0.387$. Let's estimate the values $t_b$, $h_\eta$, $\eta$. As is known, the minimum duration of inflation is $t_{inf}\sim 10^{-37}$ sec, and it should last at least $60$ Hubble times (e-folds). The graph Fig.\ref{ParamAniz} shows that at times $h_\eta\cdot t>1$  the anisotropy $\beta'_\alpha$ is already small, therefore $a(t)\propto e^{h_\eta t}$ will be almost throughout the entire inflation time interval. Then we have $h_\eta\cdot t_{inf} \sim 60$, or $h_\eta \approx 6\cdot 10^{38}$ sec$^{-1}$. From (\ref{hett12}) it follows $|\eta|\approx 10^{-78}$ sec$^2$. The start time of inflation is $t_b\approx6.5\cdot10^{-3}\cdot t_{inf}=6.5\cdot10^{-40}$ sec $>t_P$. Planck time boundary $t_P\approx 5.4\cdot10^{-44}$ is not crossed. The period before inflation is much shorter than the inflation period: $(t_b-t_P)/t_{inf}\approx 6.5\cdot10^{-3}$. Thus, after a short post-singularity era the Universe enters a primary quasi-de Sitter epoch with the parameter $h_\eta$. NMKC in isotropic space gives inflation at the beginning of the Universe (see (\ref{00sdf})).
Anisotropy of space shifts the onset of inflation. This is obvious in the model without the magnetic field (\ref{albet1}) -- at the beginning shear scalar $\sigma_0^2 a_0^6 a^{-6}$ dominates. Inflation is shifted to time $h_\eta\cdot \tau_b\approx 0.382$. This is not much different from $h_\eta\cdot t_b\approx 0.387$. The influence of the magnetic field is leveled out during the transition period. Its density is limited at the singularity point. The applicability of the model is limited to the period before and during primary inflation $h_\eta\cdot(t_b+t_{inf}) \sim 60$\, ($t_b+t_{inf}\thicksim 10^{-37}$ sec).

\subsection{Constraints on model parameters}

The model with $l>0$  becomes almost isotropic in a short period of time ($1<h_\eta \cdot t\lesssim 2$). The model is approximately defined by the equalities (\ref{00sdf}), (\ref{0sdf}). In the isotropic cosmological model two conditions related
to scalar perturbations must be satisfied \cite{DeFelice1}:
\begin{equation}c_S^2\equiv\frac{3(2w_1^2w_2H-w^2_2w_4+4w_1w_2\dot w_1-2w_1^2\dot w_2)}{w_1(4w_1w_3+9w_2^2)}\geq 0\,,\label{cS}
\end{equation}
for the avoidance of Laplacian instabilities associated with the
scalar field propagation speed, and
\begin{equation}
Q_{S}\equiv\frac{w_{1}(4w_{1}w_{3}+9w_{2}^{2})}{3w_{2}^{2}}>0\,,
\label{QS}
\end{equation}
for the absence of ghosts. For the tensor perturbations the conditions for avoidance of ghost
and Laplacian instabilities are respectively written as
\cite{DeFelice1}
\begin{equation}Q_T\equiv\frac{w_1}{4}>0\,,\, c^2_{T}\equiv\frac{w_4}{w_1}\geq 0\,.\label{csT}
\end{equation}

For the subclass of Horndeski theory (\ref{lagr1}) we have
$$w_1=\mu+\eta X,\, w_2=2H(\mu+3\eta X),$$\begin{equation}w_3=3(\varepsilon X-3\mu H^2-18\eta XH^2),\, w_4=\mu-\eta X.\end{equation}

Using the consequences of  (\ref{00sdf}) and (\ref{0sdf}):
\begin{equation}X= \frac{3\varepsilon \mu}{2}(\varepsilon_\Lambda h^2_\Lambda-h^2_\eta),\, H= h_\eta,\end{equation}
we obtain:
$$w_1=\frac{\mu}{2}\left[1+\varepsilon_\Lambda\frac{h^2_\Lambda}{h^2_\eta}\right],\, w_2=\mu h_\eta\left[-1+3\varepsilon_\Lambda\frac{h^2_\Lambda}{h^2_\eta}\right],$$
\begin{equation}\label{w1234}w_3=\frac{9}{2}\mu h^2_\eta\left[3-5\varepsilon_\Lambda\frac{h^2_\Lambda}{h^2_\eta}\right],\, w_4=\frac{\mu}{2}\left[3-\varepsilon_\Lambda\frac{h^2_\Lambda}{h^2_\eta}\right].\end{equation}
Here we assumed (\ref{ettavar}).
Taking into account (\ref{w1234}), we rewrite the conditions to  $c^2_S$
$Q_S$, $c^2_T$ and $Q_T$ in the form
$$c^2_S=\frac{1}{3}\cdot \frac{-1+3\varepsilon_\Lambda\cdot\frac{h^2_\Lambda}{h^2_\eta}}{1+\varepsilon_\Lambda\cdot\frac{h^2_\Lambda}{h^2_\eta}}\geq0,$$
$$Q_S=6\mu\left[1+\varepsilon_\Lambda\cdot\frac{h^2_\Lambda}{h^2_\eta}\right]\cdot \left[\frac{1-\varepsilon_\Lambda\cdot\frac{h^2_\Lambda}{h^2_\eta}}{-1+3\varepsilon_\Lambda\cdot\frac{h^2_\Lambda}{h^2_\eta}}\right]^2>0,$$
\begin{equation}\label{cqts}c^2_T=\frac{3-\varepsilon_\Lambda\cdot\frac{h^2_\Lambda}{h^2_\eta}}{1+\varepsilon_\Lambda\cdot\frac{h^2_\Lambda}{h^2_\eta}}\geq0,\,\,
Q_T=\frac{\mu}{8}\left[1+\varepsilon_\Lambda\cdot\frac{h^2_\Lambda}{h^2_\eta}\right]>0.\end{equation}
Inequalities (\ref{cqts}) leads to restrictions
\begin{equation}\label{sdf3}\varepsilon_\Lambda>0 \, (\Lambda>0),\, 1/3<\frac{h^2_\Lambda}{h^2_\eta}\leq 3.\end{equation}
In case $\Lambda=0$ there is Laplace instability: $c^2_S=-1/3<0$. The parameter $\Lambda$ is an important factor in the stability of the model.

Tensor-to-scalar ratio is
\begin{equation}\label{tentoscalar} r=\frac{4Q_S}{Q_T} =192\left[\frac{1-\frac{h^2_\Lambda}{h^2_\eta}}{-1+3\frac{h^2_\Lambda}{h^2_\eta}}\right]^2.\end{equation}
Here we have taken into account $\Lambda>0$.
Сonstraints from the PLANCK observations \cite{Aghanim} at the moment estimated as $r<r_0=0.1$. Therefore, the interval (\ref{sdf3}) decreases:
\begin{equation}\label{tentoscalar11} \left[\frac{1+(r_0/192)^{1/2}}{1+3(r_0/192)^{1/2}}\right]^{1/2}<\frac{h_\Lambda}{h_\eta}<
\left[\frac{1-(r_0/192)^{1/2}}{1-3(r_0/192)^{1/2}}\right]^{1/2},\,\,\text{or}\,\, 0.978<\frac{h_\Lambda}{h_\eta}<1.024.\end{equation}
The $\Lambda$-term provides a parametric degree of freedom that allows the model to be tuned to the observed data.

We have decided on the sign of $\Lambda>0$, which gives an exact expression (from (\ref{phemlp}))
\begin{equation}\label{phemlp123}\dot{\phi}^2=3\varepsilon\mu\left[ h_\Lambda^2-h_\eta^2+(h_\eta^2+ h_\Lambda^2)\cdot s\right].\end{equation}
The values $\dot{\phi}$ lie in the interval with boundaries
$$\pm \left[3\varepsilon\mu\left\{h_\Lambda^2-h_\eta^2+(h_\eta^2+ h_\Lambda^2)\cdot s_2\right\}\right]^{1/2},$$
\begin{equation}\pm \left[3\varepsilon\mu\left\{ h_\Lambda^2-h_\eta^2\right\}\right]^{1/2}.\end{equation}
Therefore there are restrictions
\begin{equation}\label{sdfa1}\varepsilon\left\{ h_\Lambda^2-h_\eta^2+(h_\eta^2+ h_\Lambda^2)\cdot s_2\right\}>0,\,\,\varepsilon\left\{ h_\Lambda^2-h_\eta^2\right\}>0.\end{equation}

For the phantom scalar field ($\varepsilon=-1$) we get:
$\eta<0$, $h_\eta^2> \frac{1+s_2}{1-s_2}\cdot h_\Lambda^2$ $\Rightarrow$ $h_\Lambda/h_\eta<0.847$.
Inequality contradicts (\ref{tentoscalar11}). The phantom scalar field is not allowed in the presented model.
For the canonical scalar field ($\varepsilon=1$):
$\eta>0$, $h_\Lambda/h_\eta>1$ $\Rightarrow$ $\Lambda\eta/\mu>1$. Inequalities (\ref{tentoscalar11}) are clarified:
\begin{equation}\label{nj45}1<\frac{h_\Lambda}{h_\eta}<\left[\frac{1-(r_0/192)^{1/2}}{1-3(r_0/192)^{1/2}}\right]^{1/2}\approx1.024.\end{equation}
Thus, we get
$$\Lambda>0,\,\, \eta>0,\,\, \varepsilon=1,$$
\begin{equation}\label{njk2} 1<\Lambda\eta/\mu<\frac{1-(r_0/192)^{1/2}}{1-3(r_0/192)^{1/2}}\approx 1.049. \end{equation}
In this case, the inflation model does not have the pathologies of instability. From (\ref{maxmagn}), (\ref{nj45}) it follows that the magnetic field, NMKC and $\Lambda$-term are similar on the energy scale: $\mathcal{E}^{({\rm em})}_{upp}\approx\mu h_\eta^2\approx\mu h_\Lambda^2$. It is taken into account here that $l\approx 2$.

For comparison, let us consider a similar analysis for the non-magnetic model. Without the magnetic field,  inequality  (\ref{tentoscalar11}) does not change. From (\ref{sdfa1}) one inequality remains: $\varepsilon\left\{ h_\Lambda^2-h_\eta^2\right\}>0$. Then for the canonical scalar field it follows (\ref{njk2}). And the possibility for the phantom scalar field is added:  $$\Lambda>0,\, \eta<0,\, \varepsilon=-1,$$ \begin{equation}\frac{1+(r_0/192)^{1/2}}{1+3(r_0/192)^{1/2}}<\Lambda|\eta|/\mu<1, \,\, \frac{1+(r_0/192)^{1/2}}{1+3(r_0/192)^{1/2}}\approx 0.956.\end{equation}  In terms of the Hubble parameter we have: $0.978<h_\Lambda/h_\eta<1$.

Thus, the model with $l>0$ has the necessary properties: isotropization during expansion, rapid transition to inflationary expansion, absence of ghost and Laplace instabilities. The magnetic field does not cause pathologies.

\section{Case $l<0$}

\begin{figure}[h]
\includegraphics[width=10cm]{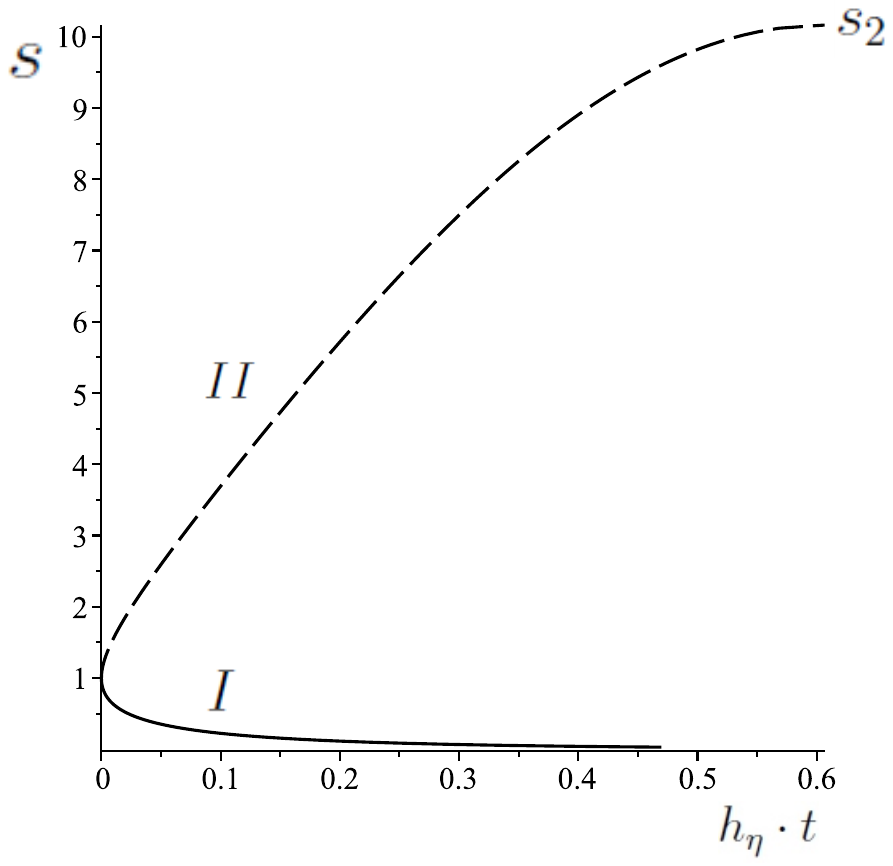}
\caption{The profile  $s(t)$. The dimensionless function $s(t)$ is associated with the magnetic field energy density $\mathcal{E}^{({\rm em})}$: $s=\frac{1}{3\mu h_\eta^2|l|}\cdot\mathcal{E}^{({\rm em})}$. In case $l<0$, the field equations have two branches of solution. The magnetic energy density is a bounded function. The branch I:  $s\in (0,1)$, the density decreases as the Universe expands. The branch II: $s\in (1,s_2]$, the density increases to a value of $s_2\approx 10.16$ at the moment of the Universe collapse. \label{Lmagnplotn}}
\end{figure}

\begin{figure}[h]
\includegraphics[width=10cm]{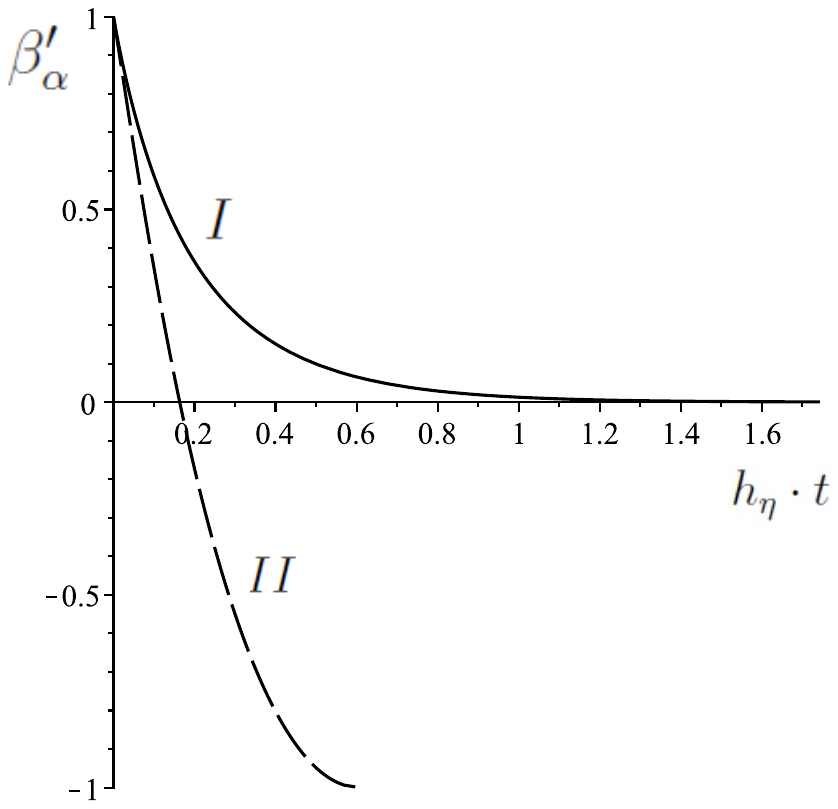}
\caption{The profile  $\beta'_\alpha$.  The branch I:  $\beta'_\alpha\in (0,1)$, the isotropization criterion is satisfied: $|\beta'_\alpha|\propto e^{-4h_\eta \cdot t}\rightarrow 0$, $h_\eta \cdot t\rightarrow +\infty$. The branch II: $\beta'_\alpha\in (-1,1)$, the isotropization process is absent. \label{Lparamaniz}}
\end{figure}

\begin{figure}[h]
\includegraphics[width=10cm]{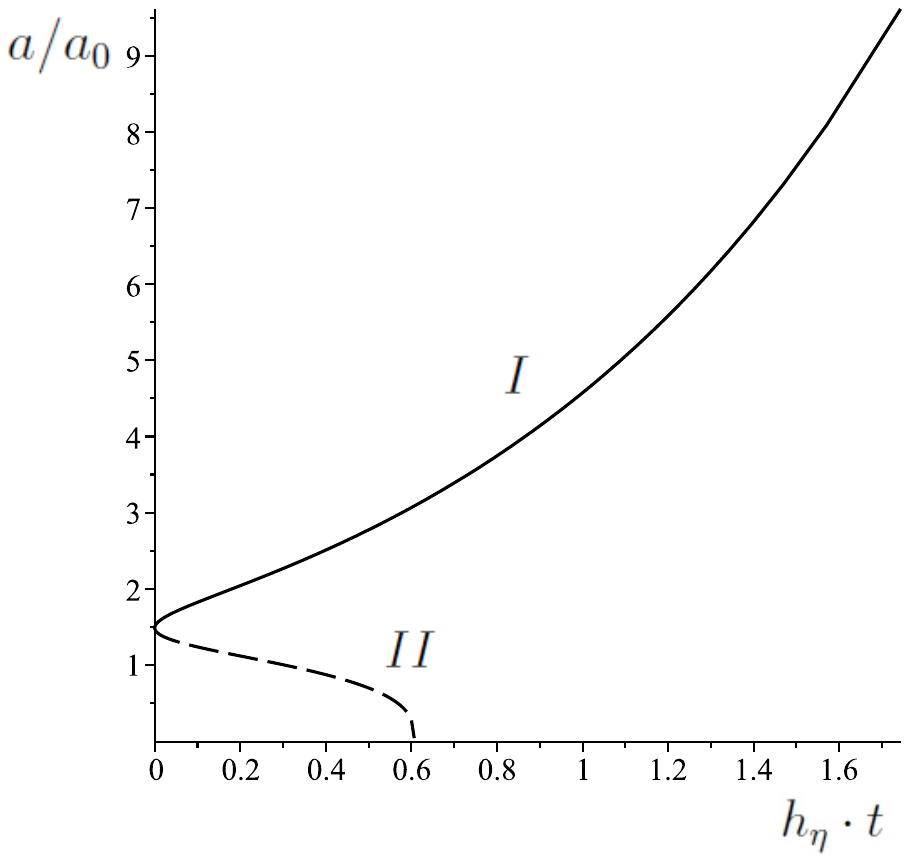}
\caption{The profile  $a(t)$. The Universe evolution begins with a non-zero value of volume. The moment $t=0$ corresponds to a cosmological singularity: $a\rightarrow \text{const}$, $\dot{\alpha}\propto 1/t^{1/2}\rightarrow\infty$, $t\rightarrow0$.  The branch I:  $a/a_0>2^{5/6}3^{-1/6}\approx 1.484$, the Universe is expanding. The branch II: $0\leq a/a_0<2^{5/6}3^{-1/6}$, the Universe that contracts in a finite time. \label{La}}
\end{figure}

Here we will consider case $\varepsilon_l=-1$ ($l<0$), i.e. $h_\eta^2+\varepsilon_\Lambda h^2_\Lambda<0$. This is possible for  $\Lambda<0$ ($\varepsilon_\Lambda=-1$).
Therefore the function $s(t)$ has values according to (\ref{s21}), where $s_2=4\cdot \sqrt{\frac{5}{3}}+5\approx 10.16$.
The value $s=1$ corresponds to the start of the count. Below in the text it will be clear why.
Let's select the "+"\, sign in (\ref{tst}). The choice of the integration constant $c=-\sqrt{\frac{5}{3}}\cdot\arcsin\left[\sqrt{\frac{3}{5}}\right]+\ln (20+4\sqrt{10})$ means the start of the count at the moment $t=0$: $s(0)=1$.
Thus
$$-4h_\eta\cdot t=-\sqrt{\frac{5}{3}}\arcsin\left[\frac{1}{4}\cdot\sqrt{\frac{3}{5}}(s-5)\right]+\ln s-$$\begin{equation}\label{Lst}
-\ln\left[5+15s+\sqrt{5}\sqrt{5+30s-3s^2}\right]-\sqrt{\frac{5}{3}}\cdot\arcsin\left[\sqrt{\frac{3}{5}}\right]+\ln (20+4\sqrt{10}).\end{equation}
The function $t(s)$ is a non-monotonic function.The graph of $s(t)$ is shown in Fig.\ref{Lmagnplotn}.
 For $s=1$ there is equality $t'_s=0$. On segment $(0,s_2)$ there are two branches of solution $s(t)$ with property (\ref{bnaal}) (see Fig.\ref{Lparamaniz}):
\begin{equation}\text{The branch I}:\,\, 0<s\leq 1,\,\, t> 0,\,\, s(+\infty)=0;\end{equation}
\begin{equation}\text{The branch II}:\,\, 1\leq s\leq s_2,\,\, 0<t\leq t_*,\,\,s(t_*)=s_2,\end{equation}
where
$$h_\eta \cdot t_*=\frac{\pi}{8}\cdot\sqrt{\frac{5}{3}}+\frac{1}{4}\cdot\sqrt{\frac{5}{3}}\cdot\arcsin\left[\sqrt{\frac{3}{5}}\right]+$$\begin{equation}+
\frac{1}{8}\ln 3-\frac{1}{4}\ln(5^{1/2}+2^{1/2})\approx 0.61.\end{equation}
To construct Fig.\ref{Lparamaniz}, a corollary of formula (\ref{sal}) was used:
\begin{equation}\label{sghwe}\dot{\beta}/\dot{\alpha}=\beta'_\alpha=-\frac{4(s^2-5s)}{s^2+10s+5}.\end{equation}
The corresponding branches of the scale factor $a(t)$ are presented in Fig. \ref{La}.
On both branches, the Universe evolution  begins with a non-zero value  $a(t)$:  $a\approx a_0\cdot2^{5/6}3^{-1/6}[1+10^{-1/4}(h_\eta\cdot t)^{1/2}]$, however $\dot{\alpha}\propto 1/t^{1/2}\rightarrow\infty$. The branch II describes the Universe that contracts in a finite time ($a/a_0\in[0,2^{5/6}3^{-1/6}]$) and the isotropization process is absent. The branch I describes the Universe that is expanding all the time, and there is the isotropization process. Next we will analyze the model I as more realistic.

For large $h_\eta\cdot t$, the magnetic field decreases according to the law $s\propto e^{-4h_\eta \cdot t}$. The model  becomes isotropic in a short period of time: $|\beta'_\alpha| \propto e^{-4h_\eta \cdot t}$ (see (\ref{sghwe}) and Fig.\ref{Lparamaniz}).  The model is approximately defined by the equalities (\ref{00sdf}), (\ref{0sdf}). The same approximation was obtained for the model with $l>0$. Hence, the conditions for the absence of pathologies are similar to (\ref{cqts}). In consequence $l<0$, three conditions are not satisfied:
$$\text{sign}(Q_S)=\text{sign}(Q_T)=\text{sign}\left[1+\varepsilon_\Lambda\cdot\frac{h^2_\Lambda}{h^2_\eta}\right]
=\text{sign}(l)=-1<0,$$
\begin{equation}c^2_T=\frac{3+\frac{h^2_\Lambda}{h^2_\eta}}{l}<0.\end{equation}
The model has ghost and Laplacian instabilities. Further exploration of the model does not make sense.

\section{Conclusion}

In this paper we have explored homogeneous
and anisotropic cosmological solutions with the magnetic field (\ref{lagrF}) in the theory of gravity
with non-minimal kinetic coupling given by the action density (\ref{lagr1}).  The matter sector is not included, since the model is studied in relation to the early times of the Universe evolution. We limit ourselves to the period before and during primary inflation.

The Horndeski theory allows anisotropy to grow over time. The question arises about isotropization. In the theory under consideration, the zero scalar charge $C_\phi=0$ imposes conditions on the anisotropy level, namely its dynamics develops in a limited region $|\dot{\beta}/\dot{\alpha}|<1$. This condition uniquely determines a viable branch of solutions of the field equations. Another consequence of $C_\phi=0$ is that inequality $\varepsilon/\eta>0$ is a necessary condition for isotropization of models, but not sufficient.

The space-time properties of NMKC without the magnetic field  coincide with the properties of the model with $L=\sqrt{-g}\left(\mu R/2-\mathbf{\Lambda}\right)$ in the Bianchi I metric, where $\mathbf{\Lambda}=\varepsilon\mu/\eta>0$. The initial parameter $\Lambda$ does not participate in the dynamics of space-time, but $\Lambda$ is included in the scalar field and it is an important factor in the stability of the model. At the beginning of the Universe, anisotropy through the shear scalar suppresses NMKC and the expansion occurs without acceleration. At time $h_\eta\cdot t\approx 0.382$ ($t\approx 6.37\cdot10^{-40}$ sec), the influence of NMKC is already significant and the expansion occurs with acceleration. The space-time becomes isotropic with time and it is de Sitter world with the parameter $h_\eta=\sqrt{\frac{\varepsilon}{3\eta}}$.

The sign of parameter $l=1+\varepsilon\eta \Lambda/\mu$ determines the properties of  cosmological models.  In studying the consequences of including the magnetic field in the model, three cases are considered:
\begin{enumerate}

  \item $l>0$. The type of singularity according to  the geometric mean $a(t)$ does not change compared to the non-magnetic model: $a\propto (h_\eta\cdot t)^{1/3} \rightarrow 0$, $\dot{\alpha}\propto 1/t\rightarrow\infty,\,\, t\rightarrow 0$. The non-magnetic model may have the thread-like singularity or the pancake singularity depending on the sign of $\dot{\beta}$. The magnetic field removes uncertainty and uniquely determines the sign $\dot{\beta}<0$. The model has  the pancake singularity: $a_{1,2}\approx\left(\frac{q^2_m}{6\mu h_\eta^2|l|s_2}\right)^{1/4}$, $a_3\approx \left(\frac{6\mu h_\eta^2|l|s_2}{ q^2_m}\right)^{1/2} \cdot b_0\cdot (h_\eta\cdot t)$. The sign of $\dot{\beta}<0$ leads to $H_{1,2}>0$, $H_3>0$, i.e. the Universe is expanding in all directions at all times. Unlike in general relativity, the magnetic field does not have a singularity. The magnetic energy density decreases from a limited value. The magnetic field does not fundamentally change the timing of the phase change of the Universe expansion ($h_\eta\cdot t\approx 0.387$ or $t\approx 6.5\cdot10^{-40}$ sec), i.e. the influence of the magnetic field is limited during the transition period. After a short post-singularity epoch, within a few $h_\eta\cdot t$ units, the anisotropy becomes small: $|\dot{\beta}/\dot{\alpha}|\propto e^{-4h_\eta \cdot t}$. The Universe enters the quasi-de Sitter epoch with the parameter $h_\eta$. Constraints on the tensor-to-scalar ratio $r<r_0=0.1$, $\dot{\phi}^2\geq 0$, the conditions for avoidance of ghost and
Laplacian instabilities lead to the inequalities: $\Lambda>0$, $\eta>0$, $\varepsilon=1$, $1<\Lambda\eta/\mu<\frac{1-(r_0/192)^{1/2}}{1-3(r_0/192)^{1/2}}\approx 1.049$.  Parameters $h_\eta$, $h_\Lambda$ are close: $1<h_\Lambda/h_\eta<1.024$. The magnetic field, NMKC and $\Lambda$-term are similar on the energy scale: $\mathcal{E}^{({\rm em})}_{upp}\approx\mu h_\eta^2\approx\mu h_\Lambda^2$. The phantom scalar field is not allowed in the presented model. For the canonical scalar field, these estimates are also valid for the non-magnetic model, but the phantom scalar field is also allowed: $\Lambda>0,\, \eta<0,\, \varepsilon=-1$, $\frac{1+(r_0/192)^{1/2}}{1+3(r_0/192)^{1/2}}<\Lambda|\eta|/\mu<1$, $\frac{1+(r_0/192)^{1/2}}{1+3(r_0/192)^{1/2}}\approx 0.956$. In terms of the Hubble parameter we have: $0.978<h_\Lambda/h_\eta<1$.

\item  In case $l=0$,  NMKC blocks the isotropization process: $|\dot{\beta}/\dot{\alpha}|\nrightarrow 0$, $h_\eta\cdot t\gg 1$.

   \item $l<0$. The type of singularity changes:  $a\approx a_0\cdot2^{5/6}3^{-1/6}[1+10^{-1/4}(h_\eta\cdot t)^{1/2}]$,  $\dot{\alpha}\propto 1/t^{1/2}\rightarrow\infty$, $t\rightarrow 0$. The Universe evolution begins with a non-zero value of volume. There are two branches  of solution:

   I. The Universe is expanding. The isotropization occurs: $|\dot{\beta}/\dot{\alpha}|\propto e^{-4h_\eta \cdot t}\rightarrow 0$, $h_\eta\cdot t\gg 1$. However, the model has ghost and Laplacian instabilities.

   II.  The Universe contracts into a point in a finite time.

 \end{enumerate}

 Thus, the inclusion of the magnetic field in NMKC leads to different possibilities. The implementation option depends on the sign $l$. The magnetic field significantly changes the effect of NMKC in case  $l\ngtr 0$, while for $l>0$ the changes are minimal. Parameters $\varepsilon$ and $\eta$ have the same sign, then the sign of $\Lambda$-term most significantly influences the sign of $l=1+\varepsilon\eta \Lambda/\mu$. Therefore, $\Lambda$-term regulates the degree of  change of the NMKC model under the influence of the magnetic field. The model with $l>0$ has the necessary properties: isotropization during expansion, rapid transition to inflationary expansion, absence of ghost and Laplace instabilities, i.e. the magnetic field does not cause pathologies. In other cases, the model has various disadvantages.

\end{document}